\documentclass[intlimits,twoside,a4paper]{article}

\usepackage{color}
\usepackage[cp1251]{inputenc}

\usepackage[eqsecnum]{cmpj3}

\issue{2018}{21}{4}{43703}
\doinumber{10.5488/CMP.21.43703}

\usepackage{bm}

\title[Generalized method of Feynman-Pines diagram technique]%
{Generalized method of Feynman-Pines diagram technique
in the theory of energy spectrum of two-level quasiparticle renormalized due to
multi-phonon processes at cryogenic temperature}
\author[M.V.~Tkach, O.Yu.~Pytiuk, O.M.~Voitsekhivska, Ju.O.~Seti]{M.V.~Tkach \footnote{E-mail: m.tkach@chnu.edu.ua},
O.Yu.~Pytiuk, O.M.~Voitsekhivska, Ju.O.~Seti}
\address{
Chernivtsi National University, 2 Kotsyubinsky St.,
58012 Chernivtsi, Ukraine}
\date{Received June 22, 2018,  in final form July 31, 2018}

\begin{document}

\maketitle

\begin{abstract}
Theory of the spectrum of localized two-level quasi-particle renormalized due to interaction with polarization phonons at cryogenic temperature is developed using the generalized method of Feynman-Pines diagram technique. Using the procedure of partial summing of infinite ranges of the main diagrams, mass operator is obtained as a compact branched chain fraction, which effectively takes into account multi-phonon processes.
It is shown that multi-phonon processes and interlevel interaction of quasiparticle and phonons cardinally change the renormalized spectrum of the system depending on the difference of energies of two states, which either resonates with phonon energy or does not. The spectrum of non-resonant system contains renormalized energies of the main states and two similar infinite series of groups of phonon satellite levels. The spectrum of a resonant system contains a renormalized ground state and infinite series of satellite groups.

\keywords diagram technique, quasi-particle, mass operator, phonon, spectrum
\pacs 71.38.-k 72.10.Di 63.20.kk
\end{abstract}

\section{Introduction}

The theory of energy spectra of quasiparticles (electrons, holes, impurity centers, excitons, etc.) interacting with phonons attracts a permanent attention due to the fact that phonons, being a dissipative sub-system at arbitrary temperature, essentially effects the physical processes and phenomena. During the starting period, the states of the systems, renormalized due to phonons, with the energies close to the energies of uncoupling quasi-particles have been intensively studied \cite{App68}. From the physical considerations it was clear that in the vicinity of so-called threshold energies (i.e., the order of quasi-particles energy plus one phonon energy) the complexes of free or bound-to-phonon states could exist. However, the lack of experimental data and mathematical problems of the theory did not contribute to the active research of even weak excited states. When a sufficient number of experimental papers on electron-phonon complexes, cyclotron resonance, impurity centers, etc. \cite{Joh66, Sum67, Lia68, Fro50, Fey62, Har67, Dea70, Rey71, Hen72} had already appeared, the theoretical investigations of the properties of the spectra of quasi-particles with the energies close to the radiation threshold of one phonon became much more active. The main theoretical results obtained during this period are presented in the review \cite{Lev74}. Here, the near-threshold phenomena are analyzed in detail within the framework of one-phonon model and approximation of vertices, defined by the diagrams of mass operator, which contains dangerous crossings only over one-quasiparticle and one-phonon lines.

The approximate methods of quantum field theory which are applied even in a ``schematic approach'' to calculate the vertices \cite{Lev74}, revealed three types of states for the impurity centers interacting with polarization phonons. The polaron states with a small effective number of phonons ($\langle N \rangle$) in it can be observed far from resonance, hybrid states ($\langle N \rangle \sim1/2$), corresponding to the dielectric modes known from the experiment --- in the resonance and bound state of quasiparticle plus phonons ($\langle N \rangle\sim 1$) --- can be observed near the resonance.

Usually, studying the spectra of quasi-particles interacting with phonons in the interval of energies that do not significantly exceed the threshold of phonon radiation, it was enough to use a one-phonon approximation. The poorly substantiated ``schematic'' approaches taking account of multi-phonon processes were used only in certain cases. In the systems with intermediate and strong coupling and in a wide interval of energies, the above mentioned processes play a dominant role. This requires the application of powerful methods of quantum field theory, such as diagonalyzing the Hamiltonians using the exact and approximated unitary transformations, Feynman and Matsubara diagram technique and so on \cite{Tkac15, Dav76, Abr12, Tka84}. Within these methods, the problem of the spectrum of one-level localized quasi-particle interacting with polarization phonons at $T=0$ K was solved exactly. It was shown that this spectrum contained the ground state shifted into the low-energy region at a quantity proportional to the coupling constant and infinite number of excited states with equidistant energies (of one-phonon energy order) in the high-energy region.

The same problem at $T\neq0$ K could not be solved. Only at an additional condition, when the probability of quasiparticle location in an arbitrary state was either zero or one, the energy spectrum of the system was obtained \cite{Dav76}. It did not depend on temperature at all and contained a ground state which was shifted into the low-energy region and equidistant levels of bound states in high- and low-energy regions. For physical reasons, this result is a strange artifact of the theoretical model with an additional condition.

For a long time the problem remained unsolved but recently there was proposed an approach \cite{Tka16}, which, using the modified method of Feynman-Pines diagram technique without artificial additional conditions, allowed taking into account not only virtual but also ``real'' multi-phonon processes. It is revealed that the complete renormalized spectrum of the system depends on temperature within the averaged phonon occupation numbers. Besides the ground state and quasi-equidistant satellite states, it contains new non-equidistant bound states of quasiparticle with many phonons. The idea of a correct method of solution appeared at the base of the known new approaches, which were proposed and used in the theory of polarons and  high-temperature superconductivity \cite{Mis09, Dev09, Mis00, Mis03, Mar10, Ebr12, Fil12, Gou16, Gou17,Pro08, Mis18}. The diagrammatic Monte Carlo method and stochastic optimization (DMC) \cite{Mis09, Dev09, Mis00, Mis03, Mar10, Ebr12, Fil12, Gou16, Gou17,Pro08, Mis18} and bold diagrammatic Monte Carlo (BDMC) \cite{Mis14, Mar17} played an important role in investigation of high excited states and revealed the reasons of some questionable results obtained for the electron-phonon interaction (for example, in the concept of a relaxed excited state). These methods are essentially based on a computer algorithm of the calculation of high-order diagrams of mass operators of Matsubara Green's functions. Therefore, one should expect that in the theory of renormalized spectrum of localized quasiparticles interacting with polarization phonons at $T\neq0$ K, there must exist the computer algorithm for constructing Feynman diagrams of mass operator and their analytical expressions of such high orders that allow a partial summing of its main continuous series. Such a computer program was presented in paper \cite{Tka16}.

As can be seen, the development of the theory of interaction of quasi-particles with phonons was almost always performed on the basis of a single-band model. However, recently there appeared an information which proves that the physics of multi-band models that describe the interacting electron-phonon systems is significantly different from single-band models, \cite{Mar17, Mol16}.

The rapid development of experimental and theoretical nano-physics during the last decade stimulated a scrupulous study of low-dimensional heterostructures, consisting of spatially confined structures (zero-dimensional quantum dots, one-dimensional quantum wires, two-dimensional quantum layers) embedded into the external bulk medium. A strong space confinement causes the quantization of quasiparticle (electrons, holes, excitons, impurities, etc.) spectra and the appearance of new types of phonons --- confined, half-space, interface, and propagating ones \cite{Tka00, Har16, Sak18, Str01}.

At first, scientists studied the properties of quasiparticles and phonons separately and then they took into account their interaction. The theory of quantum transitions between two electron bands under the effect of electromagnetic field and considering the interaction with phonons was mostly developed in one-phonon approximation without taking account of inter-level (inter-band) interaction \cite{Zha15, Tka15}.

Even the appearance of the unique devices, such as quantum cascade lasers (QCL) and quantum cascade detectors (QCD) whose main operating elements are the cascades of multi-layered quantum wells, where the relaxation of electron energy occurs due to the multi-phonon processes,  did not essentially stimulate the development of the theory of electron-phonon interaction in multi-band models. It was considered that an especially important relaxation of the electron energy in QCD was effectively provided by the radiation of single phonons. They accompanied the ``skip'' of electrons from the upper level of the active region of the previous cascade (through the ``phonon ladder'' of the extractor) to the lower level of the active region of the next cascade \cite{Fai13, Gio09}. Of course, this process plays an important role, but it is not the only one, since, as it was recently shown  in experimental studies \cite{Sak12, Sak13, Bee13}, QCD effectively operates with the ``torn phonon ladders'', where there is not a required number of equidistant electron levels to create a ``complete phonon ladder'' in extractor wells.

One of the mechanisms that can complement the ``torn phonon ladder'' of the extractor may be the quantum transitions through the high excited satellite states of the electron-phonon system of a cascade. In order to find out whether such a mechanism could ensure a successful operation of QCD with the ``torn phonon ladder'', one should theoretically study the quantum transitions between the states of the system renormalized due to electron-phonon interaction under the action of an electromagnetic field using the multi-band model and taking into account the multi-phonon processes. As far as we know, this problem is not solved yet. Moreover, due to the well-known mathematical difficulties in quantum field theory, a consistent theory of the renormalized energy spectrum of such a system has not yet been established.

In this study, we propose the generalized Feynman-Pines diagram technique, which makes possible the correct accounting of multi-phonon processes in the calculation of the renormalized energy spectrum of a two-level localized quasiparticle interacting with dispersionless phonons at $T=0$ K. As will be seen, the applied approach, firstly, will provide new interesting data on the properties of the renormalized spectrum of the main and satellite states of the system, and, secondly, it can be extended to the systems of multi-band quasi-particles interacting with phonons.

The rest of the paper is organized as follows: in section~\ref{sec2}, the Hamiltonian of the system is introduced and the rules of generalized Feynman-Pines diagram technique are formulated for the matrix of mass operator (MO) of quasiparticle Green's function. In section~\ref{sec3}, the principle of a sequential partial summing of the MO diagrams and its representation in the form of a infinite branched chain fraction is proposed. In section~\ref{sec4}, this method is used for the calculation of the renormalized spectrum of ground state and phonon complexes of satellite states of the system. The obtained results and conclusions are presented in section~\ref{sec5}.
\vspace{-2mm}

\section{Hamiltonian of the system. Mass operator of quasiparticle Green's function at $T=0$ K}\label{sec2}

\vspace{-1mm}
The system of localized two-level quasiparticle interacting with polarization phonons is described by Hamiltonian of Fr\"ohlich type, like in \cite{Fro50, Tkac15, Dav76}
\begin{equation} \label{GrindEQ__1_}
\widehat{H}=\sum _{\mu =1}^{2}E_{\mu } a_{\mu }^{+}  a_{\mu }^{} +\sum _{\vec{q}}\Omega {\rm \; }\Big(b_{\vec{q}}^{+} b_{\vec{q}}^{} +\frac{1}{2} \Big) +\sum _{\vec{q}}\sum _{\mu _{1} ,\mu _{2} =1}^{2}\varphi _{\mu _{1} \mu _{2} }  a_{\mu _{1} }^{+} a_{\mu _{2} } \big(b_{\vec{q}}^{} +b_{-\vec{q}}^{+} \big).
\end{equation}

Here, $E_{\mu}=E+d_{\mu} $ are the energies of uncoupling quasiparticle (further, $d_{\mu=1}=0$, $d_{\mu=2}=d$), $\Omega $ is an energy of polarization phonon, $\varphi _{\mu _{1} \mu _{2} } $ are the binding constants characterizing the intra-level (at $\mu _{1} =\mu _{2} $) and inter-level (at $\mu_{1} \ne \mu _{2} $) interaction. All these quantities are assumed as known parameters of the system. The operators $a_{\mu },\,  a_{\mu }^{+} $ and $b_{\vec{q}},\, b_{\vec{q}}^{+} $ satisfy Bose commutative relationships.

The energy spectrum of the system, renormalized due to the interaction, at cryogenic temperature (formally, at $T=0$ K), is obtained using the method of Feynman-Pines diagram technique \cite{Fey62, Tkac15, Abr12} for the Fourier image of casual Green's function [$G_{\mu \mu } (\omega )$] of quasiparticle. According to the general theory~\cite{Abr12} and taking into account the Hamiltonian \eqref{GrindEQ__1_}, the matrix functions $G_{\mu \mu'} (\omega )$ satisfy the system of two equations.
\begin{equation} \label{GrindEQ__2_}
G_{\mu \mu '} (\omega )=(\omega -E_{\mu } + \rm i\eta )^{-1} \bigg[\delta _{\mu \mu'} +\sum _{\mu _{1} =1}^{2}M_{\mu \mu _{1} }  (\omega )G_{\mu _{1} \mu '} (\omega )\bigg],\qquad \mu , \mu'=1, 2; \qquad (\eta \to +\infty )\,,
\vspace{-1mm}
\end{equation}
where $M_{\mu \mu _{1} } (\omega )$ is a matrix MO.
\newpage

The further calculations are performed using the dimensionless functions, variables and constants
\begin{equation} \label{GrindEQ__3_}
\begin{array}{l} {g_{\mu \mu '} (\xi )=G_{\mu \mu '}  \Omega ;\qquad {\Large{\textbf{m}}}_{\mu \mu'}(\xi )=M_{\mu \mu '} \Omega^{-1}; \qquad \xi =(\omega -E)\Omega ^{-1} }; \\ {\xi _{\mu } =(\omega -E_{\mu } )\Omega ^{-1}; \qquad \alpha _{\mu _{1} \mu _{2} } =\varphi _{\mu _{1} \mu _{2} } \Omega ^{-1}; \qquad \quad \delta =d \frac{}{} \Omega ^{-1} .} \end{array}
\end{equation}

Then, the exact solution of equation \eqref{GrindEQ__2_} for the dimensionless function of $m$-th level $g_{\mu }(\xi )=g_{\mu \mu }(\xi )$ is defined by Dyson equation ($\hbar=1$)
\begin{equation} \label{GrindEQ__4_}
g_{\mu }(\xi )=\big[ \xi_{\mu }-{\Large{\textbf{m}}}_{\mu }(\xi )\big]^{-1}
\end{equation}
through the complete MO ${\Large{\textbf{m}}}_{\mu } (\xi )$ of this level, which can be written as a sum of two components
\begin{equation} \label{GrindEQ__5_}
{\Large{\textbf{m}}}_{\mu } (\xi )={\Large{\textbf{m}}}_{\mu \mu } (\xi )+ {\Large{\textbf{m}}}_{\mu \mu }^{g} (\xi ).
\end{equation}
Here, the component ${\Large{\textbf{m}}}_{\mu \mu } (\xi )$ is determined only by the diagonal matrix elements of the complete MO ${\Large{\textbf{m}}}_{\mu \mu '} (\xi )$, while the component ${\Large{\textbf{m}}}_{\mu \mu'}^{g} (\xi )$ also contains non-diagonal components $\big[ {\Large{\textbf{m}}}_{12} (\xi ), {\Large{\textbf{m}}}_{21} (\xi )\big] $ of the same MO, which appear due to the non-diagonal elements of $g_{\mu \mu '} (\xi )$ matrix. Hence, the components ${\Large{\textbf{m}}}_{11}^{g} (\xi )$ and ${\Large{\textbf{m}}}_{22}^{g} (\xi )$ are written as follows:
\begin{equation} \label{GrindEQ__6_}
{\Large{\textbf{m}}}_{11}^{g} (\xi )=\frac{{\Large{\textbf{m}}}_{12} (\xi ){\Large{\textbf{m}}}_{21} (\xi )}{\xi _{2} - {\Large{\textbf{m}}}_{22} (\xi )} \,,\qquad {\Large{\textbf{m}}}_{22}^{g} (\xi )=\frac{{\Large{\textbf{m}}}_{12} (\xi ){\Large{\textbf{m}}}_{21} (\xi )}{\xi _{1} -{\Large{\textbf{m}}}_{11} (\xi )}\,.
\end{equation}

The matrix ${\Large{\textbf{m}}}_{\mu \mu '} (\xi )$ of the complete MO is defined by the rules of Feynman-Pines diagram technique~\cite{Abr12}, which are generalized for the case of multi-level systems. The energies in Hamiltonian~\eqref{GrindEQ__1_} are dispersionless, contrary to the classic Fr\"ohlich one-band Hamiltonian \cite{Fro50, Tkac15, Abr12}, where the energy of uncoupling quasiparticle is a function of quasi-momentum and, consequently, its MO contains diagrams both without and with the crossing phonon lines. Thus, the equivalent diagrams with and without crossing phonon lines correspond to the identical analytical expressions. Hence, in dimensionless variables~\eqref{GrindEQ__3_}, MO ${\Large{\textbf{m}}}_{\mu \mu '} (\xi )$ is obtained in such diagrammatic form which contains all possible non-equivalent diagrams without crossing phonon lines. The number of equivalent diagrams of this type is given by the integers before the respective diagrams.
\begin{equation} \label{GrindEQ__7_a}
\hfil\includegraphics[width=4.5in]{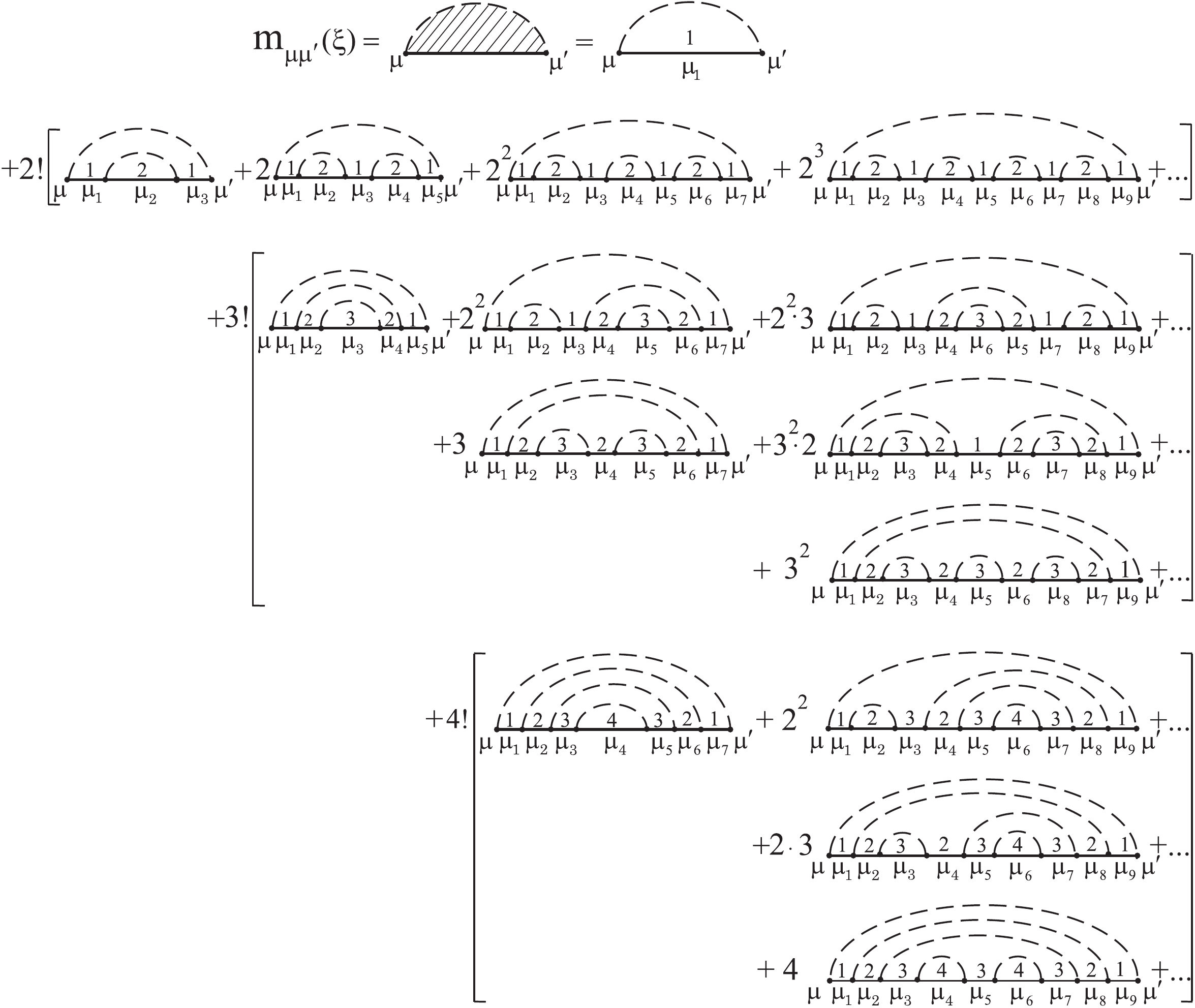} \nonumber
\end{equation}
\begin{equation} \label{GrindEQ__7_b}
\raisebox{0pt}{\hfil\includegraphics[width=4.5in]{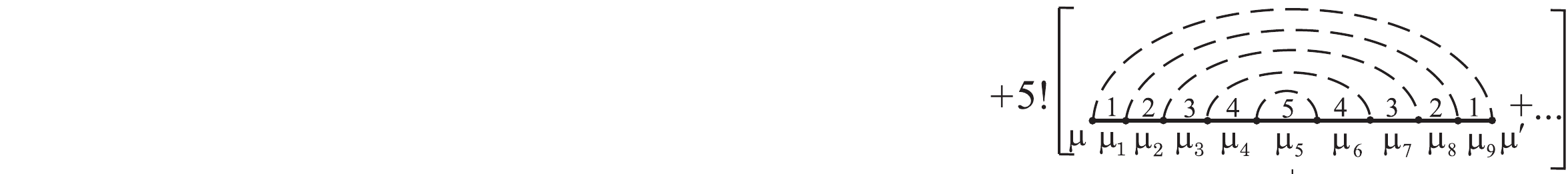}}\raisebox{12pt}{$\hspace{1mm}.$} 
\end{equation}

All MO diagrams in an arbitrary $p$-th order over the power of pairs of coupling constants, which are the same as the number of phonon lines in a respective diagram, are definitely calculated using a computer program. Further on it will become clear that in order to establish the exact rules of partial summing of infinite ranges of diagrams, it is quite sufficient to take into account all the first 64 diagrams to the seventh order inclusive.

The analytical expression for the arbitrary MO diagram is obtained as a sum over all inner indices ($\mu_{1}, \mu_{2},\dots , \mu_{N}$), except the outer indices ($\mu, \mu'$), of the products of all vertices and solid lines
\begin{equation} \label{GrindEQ__8_}
\hfil\includegraphics[width=3.2in]{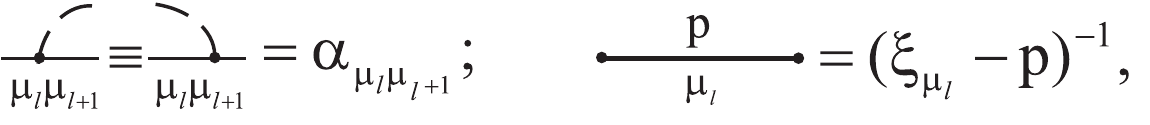}
\end{equation}
where $p$ is a number of dashed (phonon) lines placed above the solid (quasiparticle) line with index $\mu_{\textit{l}}$. For example, the first two components of ${\Large{\textbf{m}}}_{\mu \mu '} (\xi )$ have the following diagrammatic and analytical forms
\begin{equation} \label{GrindEQ__9_}
\hfil\includegraphics[width=2.3in]{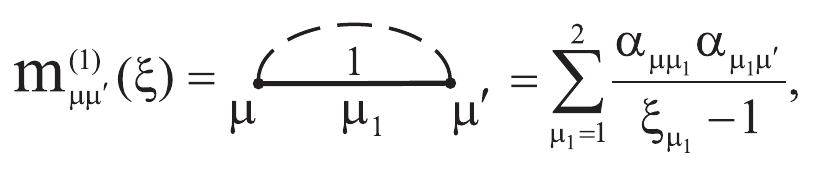}
\end{equation}
\begin{equation} \label{GrindEQ__10_}
\hfil\includegraphics[width=4.0in]{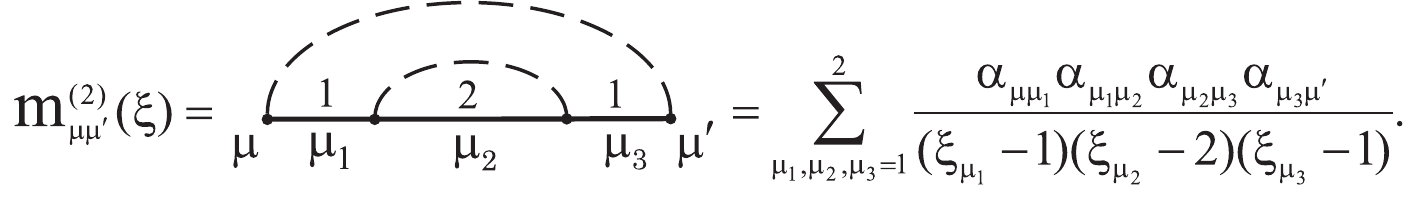}
\end{equation}

\section{Partial summing of MO diagrams}\label{sec3}

For partial summing of diagrams in the complete MO ${\Large{\textbf{m}}}_{\mu \mu '} (\xi )$, it is convenient to group them into classes which are separated by brackets in expression \eqref{GrindEQ__7_b}. It is clear that the $p$-th class of diagrams together with the factor $p!$ contains an infinite number of only exactly those diagrams (together with numerical factors), whose arbitrary blocks, in their turn, contain not more than p dashed lines over any solid line. Such a class of diagrams is further referred to as a partial $p$-phonon MO and is denoted as ${\Large{\textbf{m}}}_{\mu \mu '}^{[p]} (\xi )$. For example, the partial 2-phonon MO is
\begin{equation} \label{GrindEQ__11_}
\raisebox{0pt}{\hfil\includegraphics[width=3.3in]{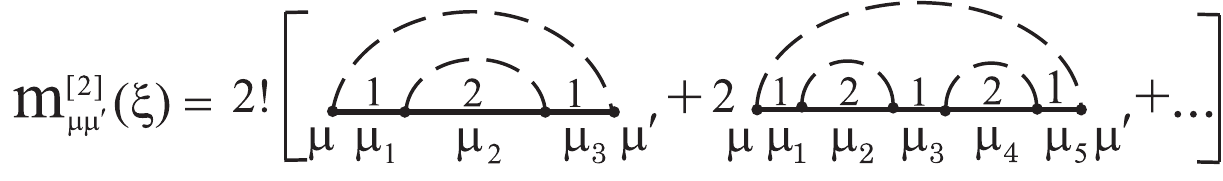}}\raisebox{12pt}{$\hspace{1mm}.$}
\end{equation}
Then, the complete MO ${\Large{\textbf{m}}}_{\mu \mu'} (\xi )$ can be written as follows:
\begin{equation} \label{GrindEQ__12_}
{\Large{\textbf{m}}}_{\mu \mu '} (\xi )={\Large{\textbf{m}}}_{\mu \mu '}^{(1)} (\xi )+\sum _{p=2}^{\infty }{\Large{\textbf{m}}}_{\mu \mu '}^{[p]} (\xi )\,.
\end{equation}

In each diagram of the MO, the analytical contributions of its elements are summed up over all $\mu_{l}$ indices, since, in general form, none of them  is multiplicative. However, the analytical expressions show that each diagram can be expressed as a sum of two components: multiplicative (m) and non-multiplicative (nm) one. For this sake, in any diagram, in which summing is performed over all inner indices (for example, $\mu_{1}$, $\mu_{2}$, $\mu_{3}$,  $\mu_{4}$,  $\mu_{5}$, $\mu_{6}$,  $\mu_{7}$) one should separate the multiplicative component with certain equal indices (for example, $\mu_{1}$, $\mu_{2}$, $\mu_{3}$, $\mu_{4}$, $\mu_{5}=\mu_{3}$, $\mu_{6}=\mu_{2}$, $\mu_{7}=\mu_{1}$), where the summing is performed only over the indices $\mu_{ 1}$, $\mu_{ 2}$, $\mu _{3}$, $\mu_{4}$, and non-multiplicative component with the indices ($\mu_{1}$, $\mu_{2}$, $\mu_{3}$, $\mu_{4}$, $\bar{\mu }_{5} =\mu _{5} \ne \mu _{3}$, $\bar{\mu }_{6} =\mu _{6} \ne \mu_{2}$, $\bar{\mu }_{7} =\mu _{7} \ne \mu _{1} $), where summing is performed over all these indices. In this approach, the diagrams of the second and third order over the number of phonon lines can be expressed in the following form:
\begin{align} \label{GrindEQ__13_}
{\Large{\textbf{m}}}_{\mu \mu'}^{(2)} (\xi )={\Large{\textbf{m}}}_{\mu \mu '}^{(2, m)} (\xi )+{\Large{\textbf{m}}}_{\mu \mu '}^{(2, nm)} (\xi ) \qquad \qquad \qquad
\nonumber\\
\hfil\includegraphics[width=3.3in]{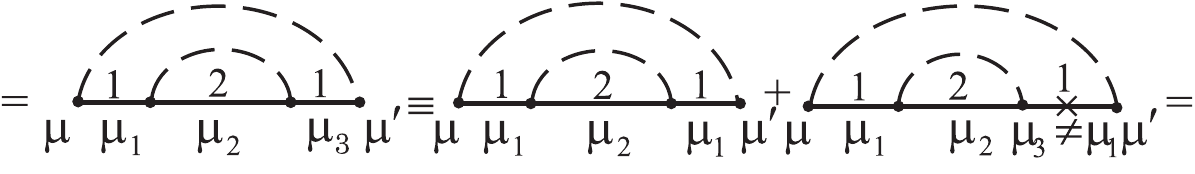}\qquad
\nonumber\\
=\sum _{\mu _{1} =1}^{2}\frac{\alpha _{\mu \mu _{1} } {\Large{\textbf{m}}}_{\mu _{1} \mu _{1} }^{(1)} (\xi -1)\, \alpha _{\mu _{1} \mu '} }{(\xi _{\mu _{1} } -1)^{2} } +\sum _{\mu _{1} =1}^{2}\, \frac{\, \alpha _{\mu \mu _{1} } }{\xi _{\mu _{1} } -1}  \sum _{\mu _{3} \ne \mu _{1} }^{2}\frac{{\Large{\textbf{m}}}_{\mu _{1} \mu _{3} }^{(1)} (\xi -1) \alpha _{\mu _{3} \mu'} }{\xi _{\mu _{3} }-1}\,,
\end{align}
\begin{equation} \label{NEW_3_4_EPS}
\hfil\includegraphics[width=5.7in]{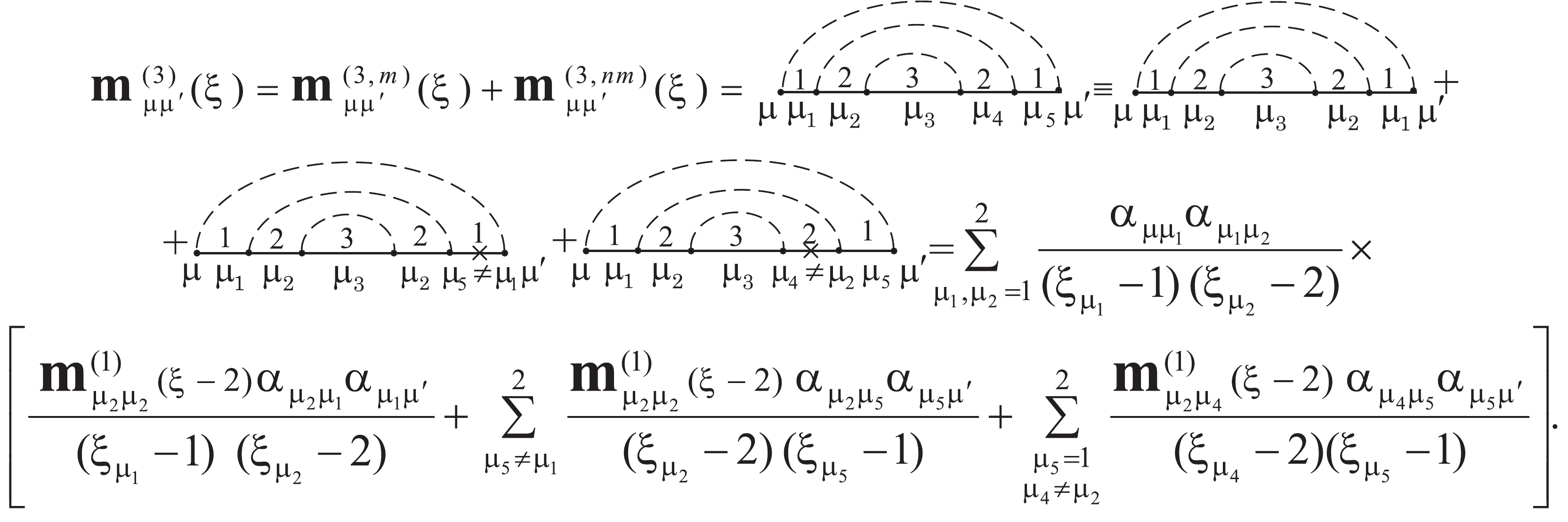}
\end{equation}

Separating in complete MO \eqref{GrindEQ__7_b} all diagrams of all orders at m- and nm-classes and taking into account that one-phonon MO ${\Large{\textbf{m}}}_{\mu \mu '}^{(1)} (\xi )$ belongs to the m-class of diagrams, from \eqref{GrindEQ__12_} we obtain:

\begin{equation} \label{GrindEQ__15_}
{\Large{\textbf{m}}}_{\mu \mu '} (\xi )={\Large{\textbf{m}}}_{\mu \mu '}^{[m]} (\xi )+{\Large{\textbf{m}}}_{\mu \mu '}^{[nm]} (\xi )=\sum _{p=1}^{\infty }{\Large{\textbf{m}}}_{\mu \mu '}^{[p, m]} (\xi ) +\sum _{p=2}^{\infty }{\Large{\textbf{m}}}_{\mu \mu '}^{[p,\, nm]} (\xi ) .
\end{equation}
For the system of a two-level quasiparticle that weakly couples with the phonons under the condition that the intra-level interaction is much bigger than the inter-level interaction ($\alpha _{11} \geqslant \alpha _{22} \gg \alpha _{12} $, that is fulfilled for the majority of semiconductor structures), in the diagrams of any $p$-th order, the multiplicative components $\big\{{\Large{\textbf{m}}}_{\mu \mu '}^{[p, m]} (\xi )\big\}$ are much bigger than non-multiplicative ones $\big\{{\Large{\textbf{m}}}_{\mu \mu '}^{[p,\, nm]} (\xi )\big\}$. Taking into account this condition, the non-multiplicative MO components are not considered further. However, we should note that for the systems with an arbitrary intralevel and interlevel interactions, non-multiplicative MO components must be taken into account either as corrections, or by separating the new types of multiplicative diagrams from them, followed by their partial summing. The component ${\Large{\textbf{m}}}_{\mu \mu '}^{(m)} (\xi )$, due to its multiplicative structure, can be exactly partially summed. Finally, it assumes a form of an infinite branched chain fraction with simple typical links.

A complete partial summing of all diagrams of ${\Large{\textbf{m}}}_{\mu \mu '}^{(m)} (\xi )$ MO is performed within the successive renormalization of the energy denominators of its components of a lower order by the contributions of all the diagrams of the next higher order. Thus, let us observe the diagrams of MO ${\Large{\textbf{m}}}_{\mu \mu '}^{(m)} (\xi )$, which describe one-, two- and three-phonon processes
\begin{equation} \label{NEW_3_6}
\hfil\includegraphics[width=4.6in]{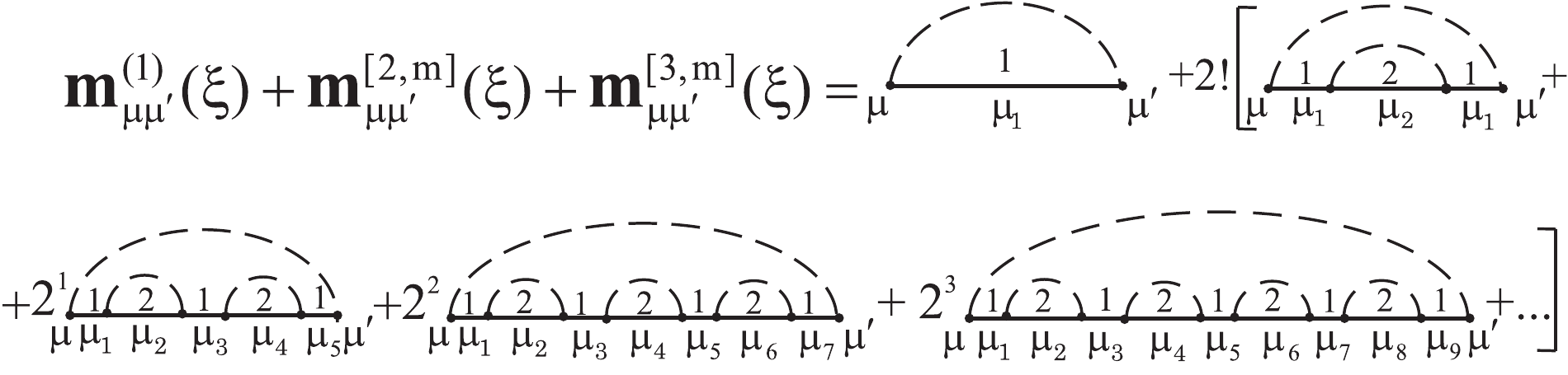} \nonumber
\end{equation}
\newpage
\vspace{-8mm}
\begin{equation} \label{GrindEQ__16_b}
\raisebox{0pt}{\hfil\includegraphics[width=4.5in]{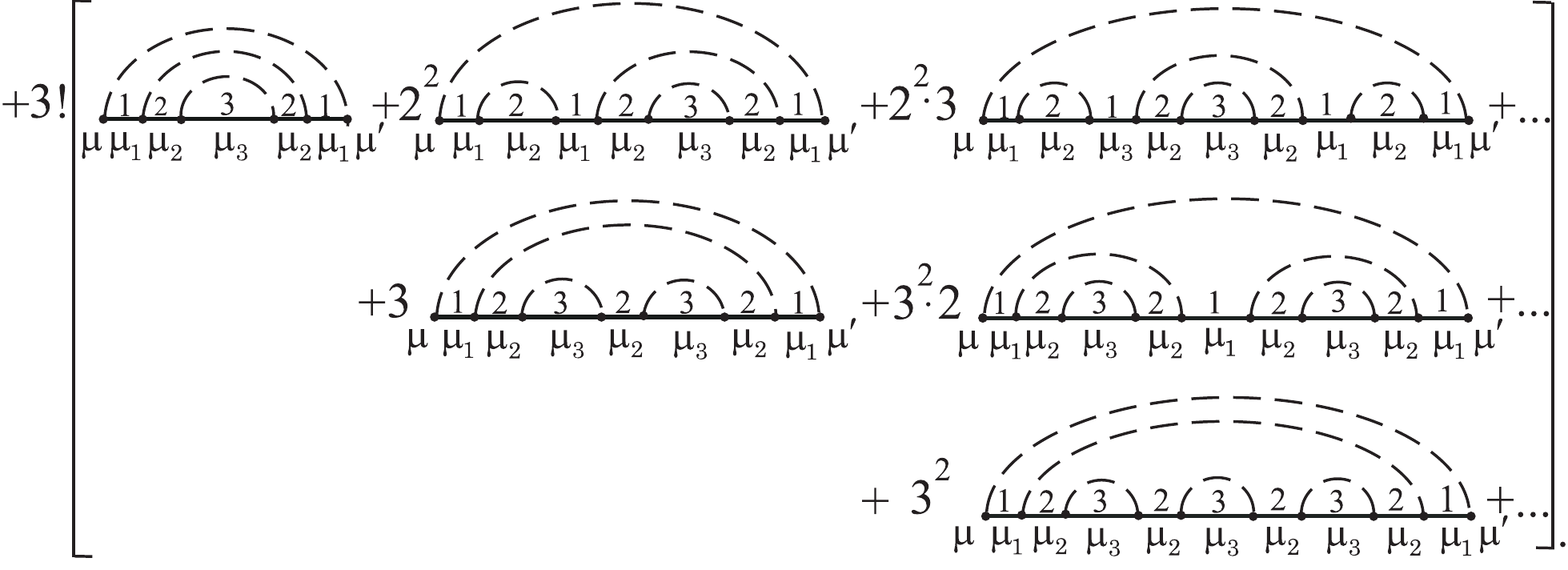}}\raisebox{93pt}{$\hspace{0mm}.$}\raisebox{4.75pt}{$\hspace{-1.65mm}\textcolor{white}{,}$}
\end{equation}

\vspace{-2mm}
Now, using the multiplicative structure and the rules of diagram technique, it is possible to perform the first stage of the renormalization of one-phonon MO by all only two-phonon processes
\begin{equation} \label{GrindEQ__17_}
\hfil\includegraphics[width=4.3in]{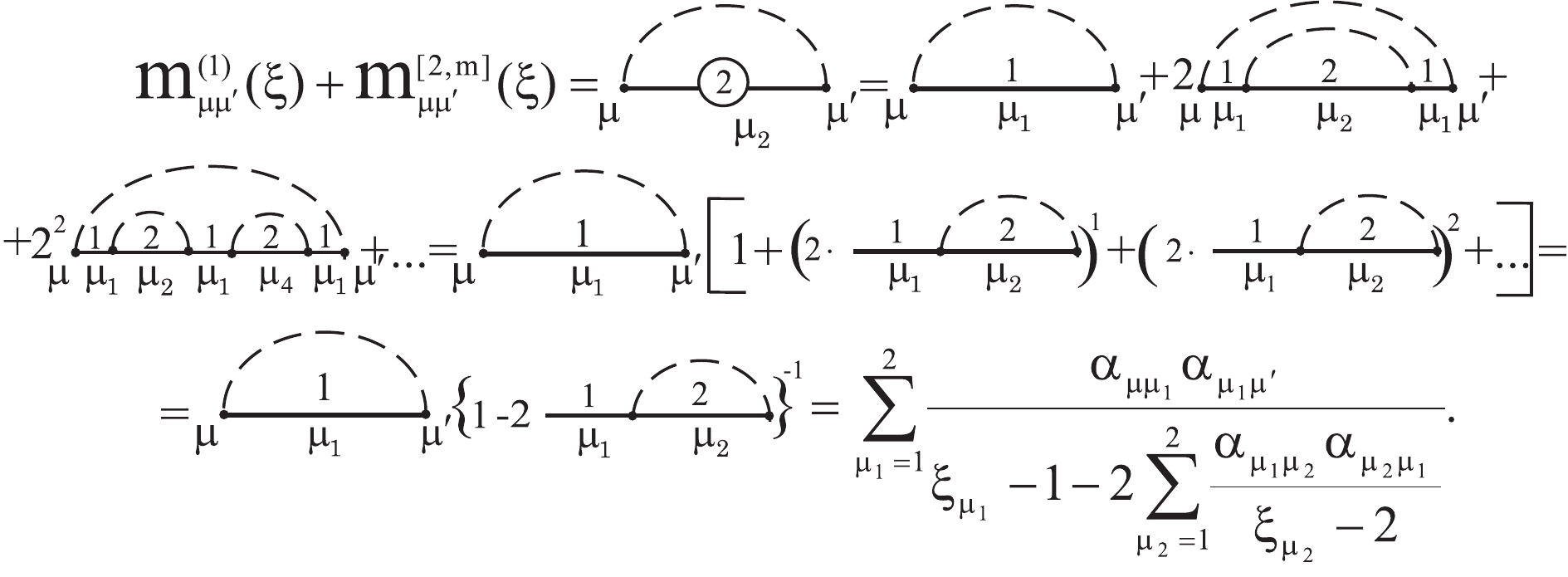}
\end{equation}

\vspace{-2mm}
In order to perform the next (second) stage, we should note that, as it is seen from formula \eqref{GrindEQ__16_b}, each two-phonon diagram, belonging to infinite series, can be renormalized by the  summing of the respective infinite series of three-phonon diagrams. In particular, the first component of two-phonon diagrams is reformed by the infinite series of three-phonon diagrams in the following way:
\begin{equation} \label{GrindEQ__18_}
\hfil\includegraphics[width=4.8in]{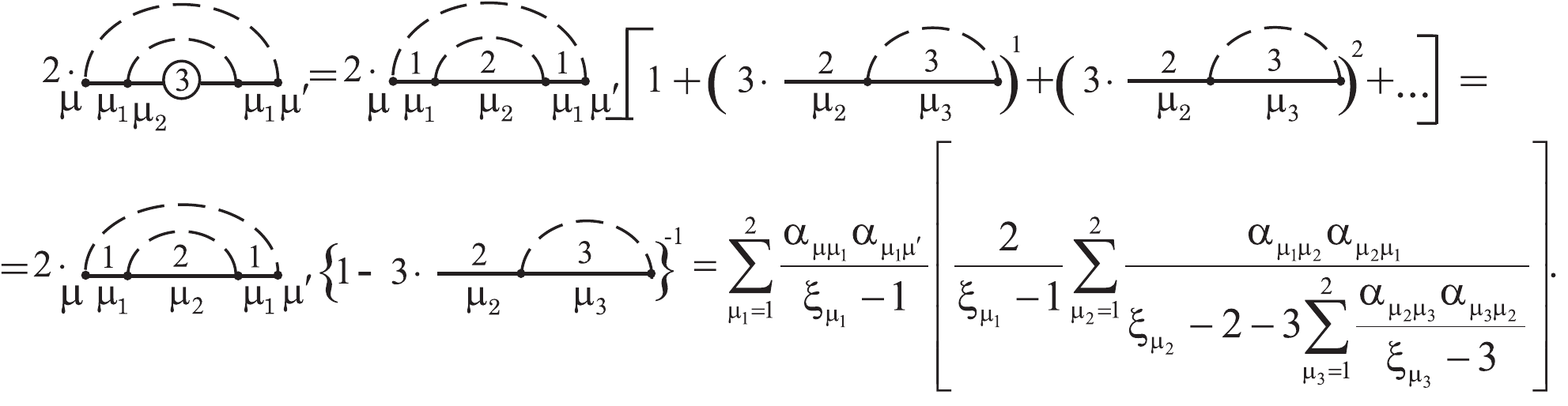}
\end{equation}
The second component of two-phonon diagrams is reformed by the following infinite series of three-phonon diagrams:
\begin{equation} \label{GrindEQ__19_}
\raisebox{0pt}{\hfil\includegraphics[width=4.8in]{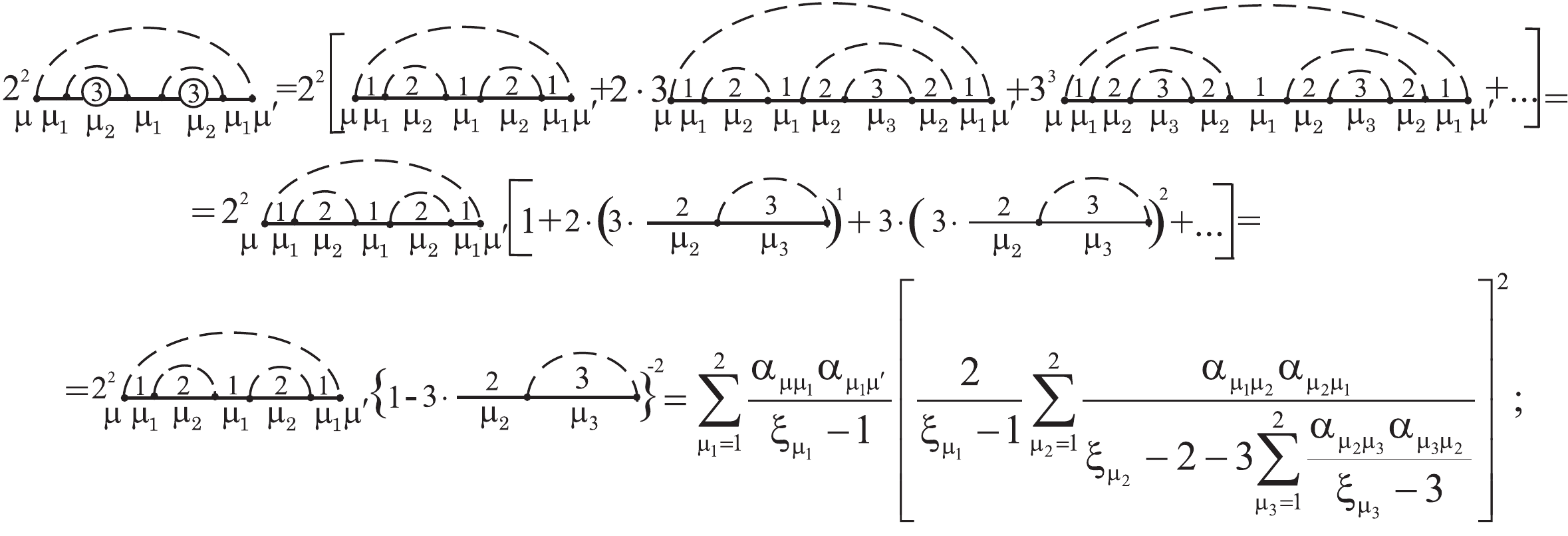}}\raisebox{26pt}{$\hspace{-4.7mm} \textcolor{white}{\blacklozenge}\hspace{-3mm}.$}
\end{equation}
In the same way, all two- and three-phonon multiplicative diagrams are summed and, taking into account the one-phonon MO, we obtain:
\begin{align} \label{GrindEQ__20_}
\hfil\includegraphics[width=4in]{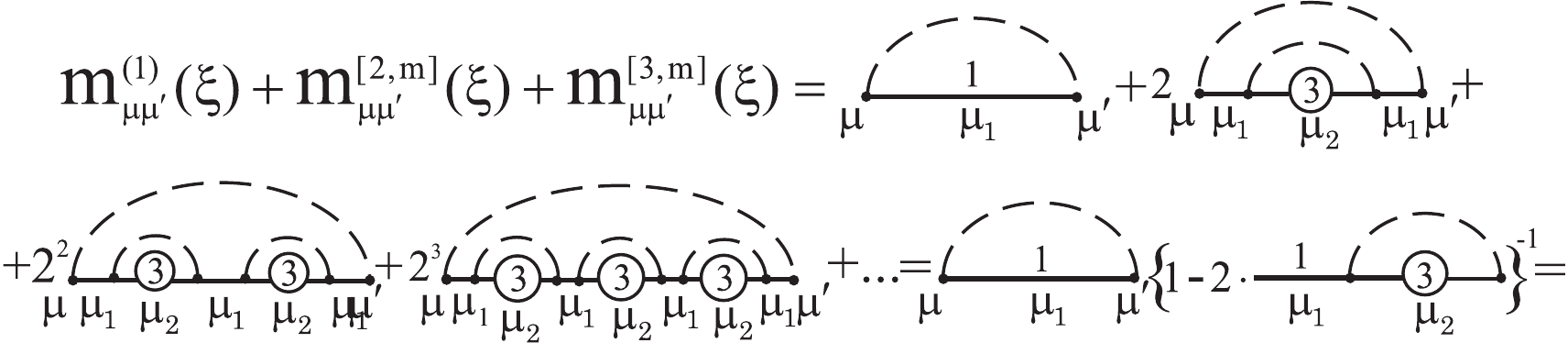}
\nonumber\\
=\sum _{\mu _{1} =1}^{2}\frac{\alpha _{\mu \mu _{1} } \alpha _{\mu _{1} \mu '} }{\xi _{\mu _{1} } -1-2\displaystyle\sum _{\mu _{2} =1}^{2}\frac{\alpha _{\mu _{1} \mu _{2} } \alpha _{\mu _{2} \mu _{1} } }{\xi _{\mu _{2} } -2-3\displaystyle\sum _{\mu _{3} =1}^{2}\frac{\alpha _{\mu _{2} \mu _{3} } \alpha _{\mu _{3} \mu _{2} } }{\xi _{\mu _{3} } -3}  }  }\,.\qquad \qquad 
\end{align}

Using the reforming of $p$-phonon multiplicative diagrams by $(p+1)$-phonon diagrams with their further summing, we obtain the exact representation of completely renormalized MO ${\Large{\textbf{m}}}_{\mu \mu '}^{(m)} (\xi )$ in the form of infinite branched chain fraction
\begin{equation} \label{GrindEQ__21_}
{\Large{\textbf{m}}}_{\mu \mu '}^{(m)} (\xi )=\sum _{\mu _{1} =1}^{2}\frac{\alpha _{\mu \mu _{1} } \alpha _{\mu _{1} \mu '} }{\xi _{\mu _{1} } -1-2\displaystyle\sum _{\mu _{2} =1}^{2}\frac{\alpha _{\mu _{1} \mu _{2} } \alpha _{\mu _{2} \mu _{1} } }{\xi _{\mu _{2} } -2-3\displaystyle\sum _{\mu _{3} =1}^{2}\frac{\alpha _{\mu _{2} \mu _{3} } \alpha _{\mu _{3} \mu _{2} } }{\xi _{\mu _{3} } -3-...-p\displaystyle\sum _{\mu _{p} =1}^{2}\frac{\alpha _{\mu _{p-1} \mu _{p} } \alpha _{\mu _{p} \mu _{p-1} } }{\xi _{\mu _{p} } -p-...}  }  }  } \,.
\end{equation}

It is worth noting that the obvious advantage of the realized representation of the MO lies in the fact that each chain has a typical compact structure and, therefore, submits to a simple computer algorithm that provides the maximum speed of numerical calculations. Besides, it is also important that this representation completely solves the ``problem of sign'' in the diagram technique because the sign-varying series are absent here.

\section{Properties of the energies of main and satellite states of two-level quasiparticle interacting with phonons}\label{sec4}

As far as the renormalized energy spectrum is defined by the poles of Fourier images of  $g_{\mu } (\xi )$ Green's functions and for the system under study, the imaginary term of MO is absent at $T=0$ K, then, according to \eqref{GrindEQ__4_}, the poles of the functions $g_{1} (\xi )$ and $g_{2} (\xi )$ are fixed by the equations
\begin{equation} \label{GrindEQ__22_}
\xi ={\Large{\textbf{m}}}_{1} (\xi )\,,\qquad \qquad  \xi =\delta +{\Large{\textbf{m}}}_{2} (\xi )\,.
\end{equation}
Their solutions give the same spectrum due to the presence of interaction between the quasiparticle and phonons.

\begin{figure}[t!]
	\centerline{\includegraphics[width=0.7\textwidth]{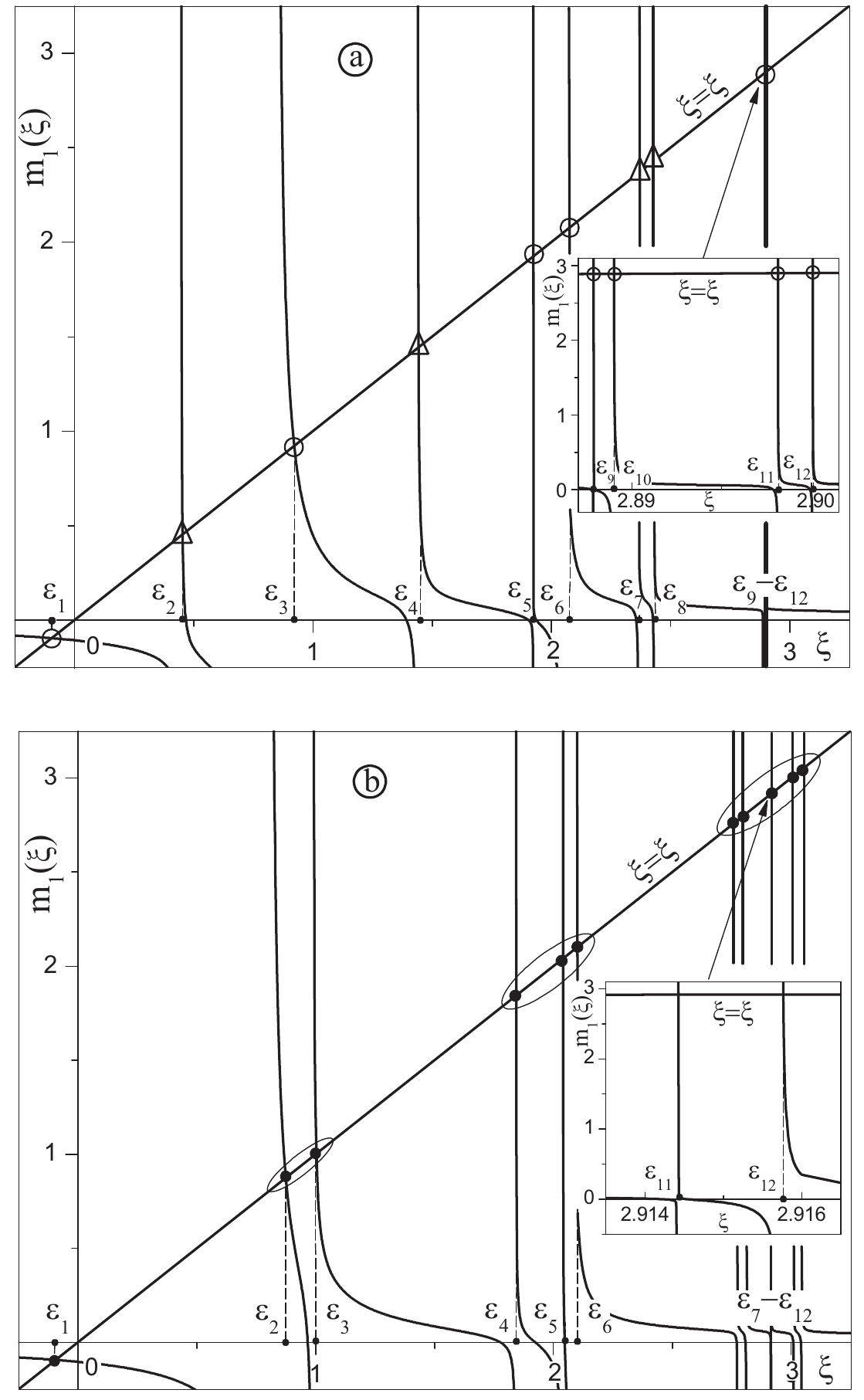}}
	\caption{Mass operator ${\Large{\textbf{m}}}_{1}^{(m)} $ as function of x  and renormalized energy spectrum of two main states ($\varepsilon _{1}, \varepsilon _{2} $) and complexes of bound-to-phonons satellite states ($\varepsilon _{3} ...\varepsilon _{12} $) in NR-case at $\delta$=0.5 (a) and in R-case at $\delta$=1 (b) at $\alpha _{11} =0.3; \alpha _{22} =0.2; \alpha _{12} =0.075$.} \label{fig-smp1}
\end{figure}

An example of functions ${\Large{\textbf{m}}}_{1} (\xi)$ and the formation of the lower part of the spectrum for the system with the coupling constants $\alpha _{11} =0.3;  \alpha _{22} =0.2;  \alpha _{12} =0.075$ in the interval of energies $\xi\leqslant 3.5$ for the non-resonant (NR) case ($\delta$=0.5) and for first resonant (R) case ($\delta$=1) are presented in figure~\ref{fig-smp1}. We should note that here and further ${\Large{\textbf{m}}}_{\mu =1} (\xi )$ \eqref{GrindEQ__5_} contains only the basic approximation where in the chain~\eqref{GrindEQ__21_} for the ${\Large{\textbf{m}}}_{1}^{(m)} (\xi )$, not more than five-six upper links were taken into account due to convergence.

Figure~\ref{fig-smp1} shows the formation of an energy spectrum of the main ($\varepsilon _{1}, \varepsilon _{2} $) and the complexes ($\varepsilon _{3} ...\varepsilon _{12} $) of satellite states, which appear due to the interaction between a two-level quasiparticle and phonons, and how it is changed depending on either the difference of energies ($\delta$) between two levels of the non-interacting quasiparticle is multiple (resonant) or non-multiple (non-resonant) with respect to the phonon energy. It is clear that in the NR-case each of the main states creates its ``own'' group of satellite states due to the complexes of quasiparticle states bound to the respective number of phonons. The energy distance between the neighbour satellite groups of each of the main states are of the order of one phonon energy. In figure~\ref{fig-smp1}~(a), the first main level and its satellites are shown with the circles while the second main level and its satellites are shown with the triangles.

In R-case, one can see a superposition of the second main state and all its satellites with all satellite states of the first main state. Hence, in the vicinity of the energies multiple to the phonon energy, an infinite number of groups of the respective number of satellite levels  with multi-anti-crossings (MACs) is observed. In figure~\ref{fig-smp1}~(b), three such groups are marked with ovals.

\begin{figure}[t!]
	\centerline{\includegraphics[width=0.75\textwidth]{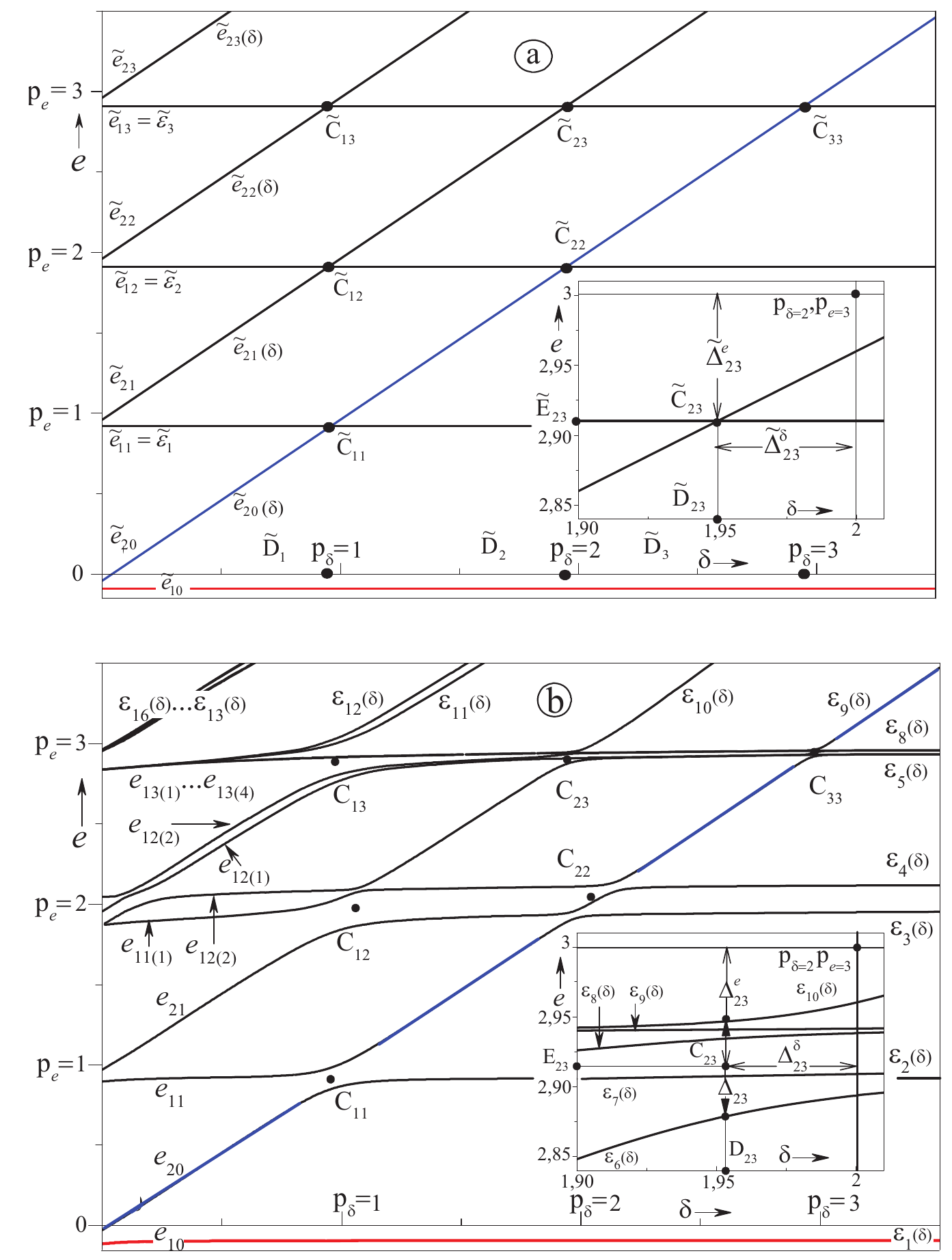}}
	\caption{(Colour online) Renormalized energy levels of the main states and complexes of satellite states of the system as functions of  $\delta $ at $\alpha _{11} =0.3, \alpha _{22} =0.2$, $\alpha _{12} =0$ (a) and $\alpha _{12} =0.075$ (b).} \label{fig-smp2}
\end{figure}

In order to uniquely classify the renormalized energy levels of the system with respect to their formation and evolution, the spectrum was calculated in a rather wide energy interval ($0\leqslant \xi \leqslant 3.5$) depending on the energy difference ($\delta$) between two levels of  the quasiparticle uncoupling with phonons. The typical example of a spectrum $\varepsilon _{n} (\delta )$ is presented in figure~\ref{fig-smp2} for the system with coupling constants $\alpha _{11} =0.3, \alpha _{22} =0.2,\alpha_{12} =0$ [figure~\ref{fig-smp2}~(a)] and $\alpha _{12} =0.075$ [figure~\ref{fig-smp2}~(b)]. These figures reveal the main properties of a renormalized energy spectrum.

The complicated non-linear dependence $\varepsilon_{n} (\delta)$ is qualitatively clear from physical considerations because it arises precisely due to the inter-level interaction between a quasiparticle and phonons ($\alpha _{12} \ne 0$). This leads not only to the shift of both main levels and their phonon satellites but also to the creation of two infinite series [in figure~\ref{fig-smp2}~(b) their number is limited by the energy interval in which the calculation was performed] of satellite groups with a different number of levels in NR- and R-regions.

Really, if the inter-level interaction is absent ($\alpha _{12} =0$), an idealized picture of the renormalized spectrum of the same system shows, figure~\ref{fig-smp2}~(a), that near each of the two main levels ($\widetilde{e}_{10} =-\alpha _{11}^{2}$, $\widetilde{e}_{20}=\delta -\alpha _{22}^{2} $) renormalized only by its ``own'' intra-level interaction with phonons, there is observed its ``own'' infinite series of equidistant satellite levels ($\widetilde{e}_{1n} =n-\alpha _{11}^{2} ,$ $\widetilde{e}_{2n}=n+\delta -\alpha _{22}^{2};  n=1, 2,\dots,\infty$). If $\delta$ increases, the behaviour of the spectrum is as follows: the whole series of the first main level and its equidistant satellite levels, shifted into the low-energy region by the quantity $\alpha_{11}^{2} $, remains  unchanged while the whole series, the second main level and its equidistant satellite levels, also shifted into the low-energy region but by the quantity $\alpha _{22}^{2} $, linearly moves into the high-energy region over~$\delta$. Consequently, in the vicinity of resonant energies ($\delta=1, 2, 3,\dots$) in the points with coordinates ($\widetilde{D}_{n} =n-\Delta =n+\alpha _{22}^{2} -\alpha _{11}^{2} ,\tilde{\varepsilon}_{n} =\tilde{e}_{1n} =n-\alpha _{11}^{2} $), the crossings of all satellite levels between each other and that of the second main level with all satellites of the first main level are observed.

Analyzing the realistic renormalized spectrum at $\alpha_{12} \ne 0$, figure~\ref{fig-smp2}~(b), it is now clear how to classify all its levels. As one can see, the realistic picture, contrary to the idealized one ($\alpha_{12} =0$), has a fundamental difference: due to an inter-level interaction between the quasiparticle and phonons, the satellite levels of both series undergo such a splitting that none of them crosses independently of $\delta $. As a result, the energies ($\varepsilon _{n=1, 2, 3,\dots} $) are single-value functions of $\delta$. Density of levels, that is, their number per unit of energy, essentially depend on the interval of $\delta$: either it is resonant ($\delta _{p=1, 2,...}\approx 1, 2, 3,\dots$) or non-resonant ($0<\delta<1$, $1<\delta<2,\dots$).

Figure~\ref{fig-smp2}~(b) shows that in all NR-intervals of $\delta$, the energy levels of all satellite states except the first level, are splitting and grouping in two series with respect to their ``own'' main levels. Like their ``own'' main levels, the respective satellite levels of the first series weakly depend on $\delta$ while those of the second series are proportional to $\delta$ in each NR-interval. In the narrow near-resonant intervals of $\delta$, due to the superposition of corresponding states, the satellite levels of both series are mixed with each other and with the second main level in such a way that, as a result, the dense groups of  MACs are formed.

In order to determine the number of split levels in both series in NR-intervals and in the groups of MACs in R-intervals, it is necessary to use two numerical indices $(p_{\delta }, p_{e} =1, 2, 3,\dots )$, which, in the units of phonon energy, fix $p_{\delta } $-th group of resonant levels and $p_{e} $-th group of phonon satellites. Then, for a further understanding of the physical origin and properties of energy spectrum, in the first NR-interval it is convenient to introduce the energies of the first $[e_{1p_{e}(k)} ]$ and second $[e_{2p_{e}(k)}]$ series of $p_{e}$-th region of satellite states, where index k numerates the split levels of this group. In order to characterize the group of levels in certain MAC, we introduce the point (center) $C_{p_{\delta}p_{e}}$ with coordinates $D_{p_{\delta}}$, $E_{p_{e} }$, which is located in the middle of its  minimal width of splitting $\Delta _{p_{\delta } p_{e} } =\min [e_{\max n} (\delta )-e_{\min n}(\delta )]$ in the vicinity of the point ($p_{\delta}, p_{e}$). Figure~\ref{fig-smp2}~(b) and the insert proves that this parameter well defines the width of ($p_{\delta } $,$p_{e} $)-th MAC along the both axes ($e$ and $\delta$). Therefore,  it is used as a characteristic size of this MAC, because beyond it, the energy levels of both series of satellites and the second main state are the linear functions of $\delta$. The proposed choice of centers and widths of MACs is based on the fact that at $\alpha_{12} \to 0$ their widths tend to zero ($\Delta _{p_{\delta } p_{e} } \to 0$), and the points $C_{p_{\delta }  p_{e} } $ are transformed into the points of crossings $\widetilde{C}_{p_{\delta } ,p_{e} } $ in figure~\ref{fig-smp2}~(a), respectively, as it should be from physical considerations. When analyzing the calculated parameters of MACs, it is more convenient not to use large values of the coordinates of their centers ($D_{p_{\delta } p_{e} } $, $E_{p_{\delta } p_{e} } $) but the respective small deviations: $\Delta _{p_{\delta } p_{e} }^{\delta } =D_{p_{\delta } p_{e} } -p_{\delta } $ and $\Delta _{p_{\delta } p_{e} }^{e} =E_{p_{\delta } p_{e} }-p_{e} $. An example of MAC parameters ($p_{\delta } =2$, $p_{e} =3$) is shown in the insert in figure~\ref{fig-smp2}~(b) and the parameters of the respective crossing are shown in the insert in figure~\ref{fig-smp2}~(a).
\newpage

The analysis of the results of calculations shows [figure~\ref{fig-smp2}~(b)] that the renormalized energy of the second main state is almost a linear function of $\delta$ in all NR-intervals between ``diagonal'' MACs ($p_{\delta }=p_{e} =p=1, 2, 3,\dots $). It separates the plane of energies ($\delta, e$) into two regions: inter-level ($0\leqslant e \leqslant \delta$) and above-level ($\delta \leqslant e \leqslant \infty$). Herein, the infinite number of satellite groups of the first series with $N_{e} =2^{p_{e} -1} $ levels is present in $p_{e} $-th group. In the above-level region, besides this series, one can see the same infinite number of groups of satellite levels of the second series with the same number of levels in $p_{e} $-th group. The arbitrary ``non-diagonal'' MAC $(p_{\delta }, p_{e}>p_{\delta })$  contains $N_{p_{\delta }, p_{e} }=2^{p_{e} -(p_{\delta } +1)} +2^{p_{e} -1} $ levels while the ``diagonal'' one $(p_{\delta }, p_{e}=p_{\delta } = p=1, 2,\dots)$ contains $N_{p} =1+2^{p-1} $ levels.

The examples of typical dependences of MAC parameters on the coupling constants describing the intra- and inter-level interaction between the quasiparticle and phonons are presented in table~\ref{tbl-smp1}. This table, like figure~\ref{fig-smp2}~(b), shows that depending on $\alpha_{12} $, the centers (coordinates of points $C_{p_{\delta }   p_{e} } $) of different MACs can be located in the plane ($\delta-e$) in different positions with respect to the points $\widetilde{C}_{p_{\delta } ,p_{e} } $ in figure~\ref{fig-smp2}~(a). However, regardless of this, an increase of $\alpha _{12} $ leads to the widening of all MACs ($\Delta _{p_{\delta } p_{5} } $). The other dependences are clearly seen in the table~\ref{tbl-smp1}.
\begin{table}[!ht]
\caption{Parameters of multi-anti-crossings ($\Delta _{p_{\delta } p_{e} } $,$\Delta _{p_{\delta } p_{e} }^{\delta } $,$\Delta _{p_{\delta } p_{e} }^{e} $) as functions of inter-level coupling constant ($\alpha _{12} $) at  $\alpha _{11} =0.3$, $\alpha _{22} =0.2$.}
\vspace{2mm}
\label{tbl-smp1}
\begin{center}
\begin{tabular}{||c|c||c|c|c||c|c||c||}  \hline \hline
$\alpha _{11} =0.3 \quad \alpha _{22} =0.2$ & $p_{\delta}p_{e}$& 1, 1 & 1, 2 & 1, 3 & 2, 2 & 2, 3 & 3, 3 \\ \hline
& $\Delta _{p_{\delta } p_{e} }^{\delta } $ & $-$0.0485 & 0.015 & $-$0.05 & 0.03 & $-$0.075 & $-$0.035 \\ \cline{2-8}
& $\Delta _{p_{\delta } p_{e} }^{e} $ & $-$0.0882 & $-$0.0525 & $-$0.1284 & 0.0099 & $-$0.0993 & $-$0.071 \\ \cline{2-8}
\raisebox{1.5ex}[0cm][0cm]{$\alpha _{12} =0.1$} & $\Delta _{p_{\delta } p_{e} } $ & 0.1611 & 0.2726 & 0.3399 & 0.1981 & 0.1122 & 0.0481 \\ \hline \hline
& $\Delta _{p_{\delta } p_{e} }^{\delta } $ & $-$0.043 & 0.061 & $-$0.025 & 0.0525 & $-$0.055 & $-$0.02 \\ \cline{2-8}
& $\Delta _{p_{\delta } p_{e} }^{e} $ & $-$0.0829 & $-$0.0210 & $-$0.1054 & 0.0217 & $-$0.0841 & $-$0.0576 \\ \cline{2-8}
\raisebox{1.5ex}[0cm][0cm]{$\alpha _{12} =0.075$} & $\Delta _{p_{\delta } p_{e} } $ & 0.1212 & 0.2283 & 0.26960 & 0.1840 & 0.0773 & 0.0279 \\ \hline \hline
& $\Delta _{p_{\delta } p_{e} }^{\delta } $ & $-$0.0395 & 0.102 & $-$0.015 & 0.065 & $-$0.035 & $-$0.01 \\ \cline{2-8}
& $\Delta _{p_{\delta } p_{e} }^{e} $ & $-$0.0793 & 0.0049 & $-$0.0934 & 0.0334 & $-$0.0694 & $-$0.0482 \\ \cline{2-8}
\raisebox{1.5ex}[0cm][0cm]{$\alpha _{12} =0.05$}& $\Delta _{p_{\delta } p_{e} } $ & 0.0810 & 0.1938 & 0.1968 & 0.1732 & 0.0483 & 0.0130 \\ \hline \hline
\end{tabular}
\end{center}
\end{table}

\section{Main results and conclusions}\label{sec5}

For a two-level localized quasiparticle interacting with polarization phonons, the Feynman-Pines diagram technique is generalized in order to calculate the renormalized spectrum of the system at $T=0$~K. A consistent and adequate method of partial summing of MO diagrams is proposed, which, avoiding the known ``problem of a sign'' in quantum field theory, effectively takes into account multi-phonon processes and presents the MO matrix in the form of a branched infinite chain fraction with the links of the same type. It provides a high speed computer calculation of renormalized spectrum.

Using Dyson equation for the system with weak quasiparticle-phonon interaction, the renormalized spectrum is obtained almost in the whole high-energy region, which contains the main levels and their infinite groups of bound-to-phonons satellite levels. It is shown that depending on whether the energy difference ($\delta$) between two levels of the uncoupling quasiparticle is resonant (multiple) to the phonon energy or non-resonant to it, the spectrum of satellite states is almost independent of $\delta$ in its interval. However, in the region of energies bigger than the energy of the second main state, the renormalized spectrum essentially depends on $\delta$. This is quite clear from physical considerations because at $T=0$~K, the multi-phonon processes can occur only with the creation of phonons, and, therefore, the bound-to-phonon states (satellites) can be formed only in the region of energies bigger than the energy of the main renormalized state. Consequently, in a two-level model, the quasiparticle in the second main state cannot create bound states with phonons in the energy interval between the two main renormalized levels. However, the inter-level interaction of both main states due to phonons causes the splitting of the $p$-th phonon satellite of the first main level (in a one-level model) into the group of $2^{p-1} $ satellite levels in a two-level model.

In the region of energy equal to and bigger than the energy of the second main level, the physical situation is different. Here, the both main states in the NR-intervals of energy create their ``own'' groups of satellite levels, and in near-R- and R-intervals they create MACs, which are groups of the resonant energy levels formed by the superposition of the second main state and satellite (bound) states of the first main state and superpositions of all satellite states created by both main states.

In general, the revealed properties of the spectrum of a two-level quasiparticle renormalized due to phonons, and, in particular, the ``ladder'' of the equidistant groups of split satellite levels present in this spectrum, can at least highlight the physical mechanism of functioning of  QCDs with cascades operating on the basis of extractors with ``torn ladders'', formed only by the electron levels of the system, in the near-infrared range. Indeed, the cascades of experimental QCD operating in middle and far infrared ranges differ from that in the near-infrared range by the fact that the sizes of quantum wells of extractors (for middle and far ranges) are selected so that they create a complete ``phonon ladder'' of equidistant (in one phonon) electron levels. In QCD operating in the near-infrared range, only the upper part of the ``phonon ladder'' can be created by selecting the sizes of deep quantum wells, while the lower part remains torn. Consequently, the QCD would not operate because the energy relaxation is impossible since the electron levels of the lower part of the ``phonon ladder'' are absent. However, it operates. Taking into account the obtained results, it seems that one of the mechanisms of relaxation of the electron energy may be not only the electron levels of the extractor but also the groups of phonon satellites located in the energy region between the two operating levels of the cascade active region.

Of course, in order to develop a consistent theory of electron-satellite mechanism of QCD operating within the torn ``phonon ladder'', the approach proposed in this paper should be generalized in such a way that it is capable of effectively taking into account the multi-phonon processes for the systems of multi-band quasiparticles interacting with polarization phonons of different modes. Such a generalization seems possible.



%
%

\ukrainianpart

\title{Узагальнений метод діаграмної техніки Фейнмана-Пайнса у теорії енергетичного спектра дворівневої квазічастинки перенормованого багатофононними процесами при кріогенній температурі}
\author{М.В.Ткач, О.Ю.Питюк, О.М.Войцехівська, Ю.О.Сеті}
\address{Чернівецький національний університет ім. Ю.Федьковича,  \\ вул. Коцюбинського, 2, 58012 Чернівці, Україна}

%
%
%

\makeukrtitle

\begin{abstract}
\tolerance=3000%
Узагальненим методом діаграмної техніки Фейнмана-Пайнса розвинена теорія перенормованого взаємодією з поляризаційними фононами спектра локалізованої дворівневої квазічастинки при кріогенній температурі. Парціальним підсумовуванням безмежних рядів основних діаграм отримано масовий оператор у компактному вигляді розгалуженого ланцюгового дробу, який ефективно враховує багатофононні процеси.
Показано, що багатофононні процеси і міжрівнева взаємодія квазічастинки з фононами кардинально змінюють перенормований спектр системи у залежності від того, чи різниця енергій між обома станами невзаємодіючої квазічастинки резонує, чи не резонує з фононною енергією. Спектр нерезонансних систем містить перенормовані енергії основних станів і дві подібні між собою безмежні серії груп фононних сателітних рівнів. Спектр резонансних систем містить перенормований основний рівень і безмежну серію сателітних груп мультиантикросингів.
\keywords діаграмна техніка, квазічастинка, масовий оператор, фонон, спектр

\end{abstract}

\end{document}